\documentclass[a4paper, 11pt]{article}

\usepackage[utf8]{inputenc}
\usepackage[T1]{fontenc}
\usepackage[english]{babel}
\usepackage{lmodern}
\usepackage{amsmath}
\usepackage{dsfont}    % Double stroke fonts
\let\mathbb\mathds

\usepackage[acronym]{glossaries}

\usepackage{graphicx}
\usepackage{psfrag}
\usepackage{subfigure}

\usepackage{enumerate}
\usepackage{multirow}

\usepackage{minibox}

\newcommand*{\ens}[1]{\mathbb{#1}}     % set
\newcommand*{\D}[1][]{\mathrm{d}{#1}}  % D of differentiation

\newcommand*{\zcrit}{z^{\mathrm{crit}}}    % critical threshold
\newcommand*{\prob}{\mathbb{P}}            % Probability
\newcommand*{\tHF}{t^{\mathrm{ref}}}       % t-High-fidelity
\newcommand*{\tLF}{t^{\mathrm{LF}}}        % t-Low-fidelity

\DeclareMathOperator{\GP}{\mathsf{G}\mathsf{P}}  % Gaussian process
\DeclareMathOperator{\N}{\mathsf{N}}             % Normal distribution
\DeclareMathOperator{\esp}{\mathbb{E}}           % Expectation
\DeclareMathOperator{\var}{\mathbb{V}ar}         % Variance
\DeclareMathOperator{\matern}{\mathcal{M}}       % Matern

\newacronym{lne}{LNE}{Laboratoire National de m\'etrologie et d'Essais}
\newacronym{l2s}{L2S}{Laboratoire des Signaux et Syst\`emes}

\newacronym{pof}{PoF}{Probability of Failure}
\newacronym{mc}{MC}{Monte-Carlo}
\newacronym{mcm}{MCM}{Monte-Carlo method}
\newacronym{gp}{GP}{Gaussian process}
\newacronym{doe}{DoE}{Design of Experiment}
\newacronym{lhs}{LHS}{Latin Hypercube Sampling}
\newacronym{nlhs}{NLHS}{Nested Latin Hypercube Sampling}
\newacronym{pdf}{pdf}{probability distribution function}
\newacronym{cdf}{cdf}{cumulative distribution function}
\newacronym{map}{MAP}{Maximum A Posteriori}
\newacronym{fb}{FB}{Fully Bayesian}
\newacronym{mf}{MF}{Multi-Fidelity}
\newacronym{sl}{SL}{Single-Level}
\newacronym{crps}{CRPS}{Continuous Ranked Probability Score}

\renewcommand\appendix{\par
  \setcounter{section}{0}
  \setcounter{subsection}{0}
  \setcounter{figure}{0}
  \setcounter{table}{0}
  \renewcommand\thesection{\Alph{section}}
  \renewcommand\thefigure{\Alph{section}\arabic{figure}}
  \renewcommand\thetable{\Alph{section}\arabic{table}}
}

\usepackage{fancyhdr}
\begin{document}

\title{Integrating hyper-parameter uncertainties\\
in a multi-fidelity Bayesian model\\
for the estimation of a probability of failure}

\author{%
  R.~Stroh$^{\dagger, \diamond, *}$, %
  J.~Bect$^{\diamond}$, %
  S.~Demeyer$^{\dagger}$, %
  N.~Fischer$^{\dagger}$ %
  and E.~Vazquez$^{\diamond}$
}

\date{\small
  $^{\dagger}\,$Mathematics and Statistics Department,\\
  \acrfull{lne},
  Trappes, France\\[1mm]
  $^{\diamond}$\acrfull{l2s},
  CentraleSup\'elec, Univ. Paris-Sud, CNRS,
  Universit\'e Paris-Saclay,
  Gif-sur-Yvette, France\\[1mm]
  $^*$E-mail: remi.stroh@lne.fr
}

\maketitle
\thispagestyle{fancy}

\begin{abstract}
A multi-fidelity simulator is a numerical model,
in which one of the inputs controls a trade-off between
the realism and the computational cost of the simulation.
Our goal is to estimate the probability of exceeding a given threshold
on a multi-fidelity stochastic simulator.
We propose a fully Bayesian approach based on Gaussian processes
to compute the posterior probability distribution of this probability.
We pay special attention to the hyper-parameters of the model.
Our methodology is illustrated on an academic example.
%We focus on the  on the prior and posterior distributions
%of the hyper-parameters of the Gaussian process
\end{abstract}

\section{Introduction}

In this article, we aim to estimate the \gls{pof} of a system described
by a multi-fidelity numerical model.
Multi-fidelity simulators are characterized by the fact
that the user has to make a trade-off between the realism of the simulation
and its computational cost,
for instance by tuning the mesh size
when the simulator is a finite difference simulator.
An expensive simulation gives a high-fidelity result,
while a cheap simulation returns a low-fidelity approximation.
A multi-fidelity approach combines different levels of
fidelity to estimate a quantity of interest.
A method for estimating probabilities of exceeding a threshold
of a stochastic multi-fidelity numerical model is proposed
in~\cite{stroh2017assessing}.
In this paper, we extend the methodology to a fully Bayesian approach.

A stochastic multi-fidelity simulator can be seen as a black-box,
which returns an output modeled by a random variable~$Z$
from a vector of inputs~$(x, t)\in \ens{X}\times \ens{R}^+$,
$\ens{X}\subset\ens{R}^d$.
The vector~$x$ is a set of input parameters of the simulation,
and the scalar~$t$ controls the fidelity of the simulation.
The fidelity increases when~$t$ decreases.
We denote by~$\prob_{x,t}$ the probability distribution of the output~$Z$ at~$(x, t)$.
We assume that an input distribution~$f_{\ens{X}}$ on the input space~$\ens{X}$
and a critical threshold~$\zcrit$ are also provided.
The \gls{pof} is the probability that the output exceeds the critical threshold
\begin{equation}
P = \int_{\ens{X}} \prob_{x,\tHF}(Z > \zcrit)f_{\ens{X}}(x)\D{x},
\label{eq:pof}
\end{equation}
where~$\tHF$ is a reference level where we would like to compute the probability.
We use a Bayesian approach based on
a multi-fidelity Gaussian process model of~$Z$ in order to
compute a posterior distribution of the \gls{pof}.
Prior distributions are added on the hyper-parameters of the Gaussian process,
so we expect that the posterior distribution of the \gls{pof} has better
predictive properties.
This approach is compared to a classical plug-in approach.

The paper is organized as follows.
Section~\ref{sec:model} explains the Bayesian multi-fidelity model.
Section~\ref{sec:inference} describes how to take into account
the hyper-parameter uncertainties to compute the posterior density of the \gls{pof}.
Section~\ref{sec:proba_fail} illustrates the methodology on an academic example.
%And Section~\ref{sec:conclusion} concludes the paper.

\section{Multi-fidelity Gaussian process}
\label{sec:model}

In this section, we present the model proposed in~\cite{stroh2017assessing}.
The output~$Z$ at~$x,t$ is assumed conditionally Gaussian
\begin{equation}
Z\vert \xi, \lambda \sim \N(\xi(x, t), \lambda(t)),
\label{eq:normalNoise}
\end{equation}
with~$\xi(x, t)$ and~$\lambda(t)$ the mean and variance functions,
the latter being assumed independent of~$x$ for simplicity.
Knowing~$\xi$ and~$\lambda$, two different runs
of the simulator produce independent outputs.
Bayesian prior models are independently added on~$\xi$ and~$\lambda$.

For the mean function~$\xi$, we use the multi-fidelity model
proposed by~\cite{picheny2013nonstationary, tuo2014surrogate}.
This model decomposes the Gaussian process~$\xi(x, t)$
in two independent Gaussian processes:
\begin{equation}
\xi(x, t) = \xi_0(x) + \epsilon(x, t),
\end{equation}
where the process~$\xi_0$ describes an ideal simulator,
which would be the result at~$t = 0$,
and~$\epsilon$ represents the numerical error of the simulator.
The model imposes~$\esp\left[\epsilon(x, 0)^2\right] = 0$.
Moreover, as the fidelity increases when~$t$ decreases,
the variance of~$\epsilon$ according to~$t$
is decreasing when~$t$ decreases.

The ideal process~$\xi_0$ is a stationary Gaussian process
with constant mean~$m$ and stationary covariance~$c_0$.
The error process~$\epsilon$ is a centered Gaussian process
with a separable covariance between~$x$ and~$t$, independent of~$\xi_0$.
Thus, the distribution of ~$\xi$ is
\begin{equation}
\xi \sim \GP(m, c_0(x - x') + r(t, t')\cdot c_{\epsilon}(x - x')).
\label{eq:gaussianProcess}
\end{equation}
The prior distribution of~$m$ is a uniform improper distribution
on~$\ens{R}$, which is a classical assumption
in ordinary kriging (see~\cite{santner2003design}).
Following the recommendations of~\cite{tuo2014surrogate},
a Mat\'ern 5/2 covariance function is selected for~$c_0$ and~$c_{\epsilon}$:
\begin{equation}
c_0(h) = \sigma_0^2 \matern_{5/2}\left(
\sqrt{\sum_{k=1}^d \left(\frac{h_k}{\rho^{0}_k}\right)^2} \right),
c_{\epsilon}(h) = \sigma_0^2 G \matern_{5/2}\left(
\sqrt{\sum_{k=1}^d\left(\frac{h_k}{\rho^{\epsilon}_k}\right)^2} \right),
\end{equation}
and a distorted Brownian covariance function for the fidelity covariance:
\begin{equation}
r(t, t') = \left(\frac{\min\left\{t, t'\right\}}{\tLF}\right)^L,
\end{equation}
with~$\sigma_0^2, G, L,
\left(\rho_k^{0}, \rho_k^{\epsilon}\right)_{1\leq k\leq d}$
$2d + 3$~positive hyper-parameters,
$\tLF$ the lowest level of fidelity (to ensure~$r(t, t') \leq 1$),
and~$\matern_{5/2}$ the covariance function~$\matern_{5/2}(h)
= \left(1 + \sqrt{5}h + \frac{5}{3}h^2\right)e^{-\sqrt{5}h}$.

In this article, even if the simulator could be observed
at any level~$t$, we assume that only $S$~levels
$t_1 > t_2 >\dots > t_S > 0$ are actually observed.
Thus, instead of inferring on the whole function~$\lambda(t)$,
we consider only the parameters~$(\lambda(t_s))_{1\leq s\leq S}$.
The vector of hyper-parameters~$\theta =
\left\{\sigma_0^2, \left(\rho^0_k\right)_{1\leq k\leq d},
G, L, \left(\rho^{\epsilon}_k\right)_{1\leq k\leq d},
\left(\lambda(t_s)\right)_{1\leq s\leq S}\right\}$ therefore has length~$2d + 3 + S$.

\section{Dealing with hyper-parameters}
\label{sec:inference}

In order to carry out a fully-Bayesian approach,
prior distributions are added on these hyper-parameters.
To simplify the estimations and the inference,
the hyper-parameters are expressed in log-scale
$l_{\theta} = \log(\theta)$,
and the joint prior distribution of~$l_{\theta}$
is chosen to be a multivariate normal distribution.
The hyper-parameters of the mean function~$\xi$ are assumed
mutually independent, and independent of the noise variance~$\lambda$.
An approximate value~$r^{\mathrm{out}}$ of the range of the output is assumed known,
and the input domain~$\ens{X}$ is assumed to be
an hyper-rectangle~$\ens{X} = \prod_{k=1}^{d}[a_k; b_k]$.
We propose, for the model described in Section~\ref{sec:model},
the following prior distributions:
\begin{subequations}
\begin{align}
l_{\sigma_0^2}& \sim
\N\left(\log\left(\frac{{r^{\mathrm{out}}}^2}{100^2}\right), \log(100)^2\right), \allowdisplaybreaks[3]\\
l_{G}& \sim
\N\left(\log\left(1\right), \log(100)^2\right), \allowdisplaybreaks[3]\\
l_{\rho_k^{0}}, l_{\rho_k^{\epsilon}} & \sim
\N\left(\log\left(\frac{b_k - a_k}{2}\right),  \log(10)^2\right),
\, 1 \leq k \leq d,\allowdisplaybreaks[3]\\
l_L & \sim \N(\log(4), \log(3)^2),\allowdisplaybreaks[3]\\
\left(l_{\lambda(t_s)}\right)_{1\leq s\leq S} & \sim
\N\left(\log\left(\frac{{r^{\mathrm{out}}}^2}{100^2}\right)\mathds{1}_S,
\log(100)^2\cdot\left((1 - c)I_S + cU_S\right)\right).
\end{align}
\end{subequations}
with~$c$ the correlation between two noise variances,
$\mathds{1}_S$ the vector of ones of length~$S$,
$I_S$ the identity matrix of size~$S$,
and~$U_S$ the square matrix of ones with size~$S$.
To select the prior distributions, we propose a reference value
for each hyper-parameter, and add a large prior uncertainty
to get weakly-informative prior distributions.
The parameters~$\sigma_0^2$ and~$G\sigma_0^2$ are assumed to be
approximatively equal to~$\left(\frac{r^{\mathrm{out}}}{100}\right)^2$.
The range parameters~$\left(\rho_k^{0}, \rho_k^{\epsilon}\right)_{1\leq k\leq d}$
are assumed to be about the half of the domain~$\rho_k \approx \frac{b_k - a_k}{2}$.
For the degree parameter~$L$, the mean is a value
recommended by~\cite{tuo2014surrogate}.

The noise variances are assumed to be about
$\left(\frac{r^{\mathrm{out}}}{100}\right)^2$,
with a large standard deviation.
However, we also assume that the prior uncertainty
on the difference between two log-noise variance are really small
with respect to the uncertainty of the noise variance,
$\var\left[\log(\lambda(t_1)) - \log(\lambda(t_2))\right]
\ll \var\left[\log(\lambda(t_1))\right]$.
Consequently, we assume~\cite{stroh2017assessing} a strong correlation
between log-noise variances, which is set to~$c = 99\%$.
This assumption helps to estimate noise variance on the levels with few observations.

Once the prior distribution are defined, we can compute
the posterior distribution conditionally to observations using Bayes theorem.
Let~$\chi_n = \left(x_i, t_i; z_i\right)_{1\leq i\leq n}$ denote
$n$ observations of the simulator.
Because of the assumption of normal output distribution
and Gaussian process with unknown mean
(Equations~\eqref{eq:normalNoise} and~\eqref{eq:gaussianProcess}), 
the prior and posterior processes conditioned by~$\theta$ are Gaussian.
Thus, for any vector of outputs~$Z$ at given input vectors,
$\pi(Z\vert \chi_n, \theta)$ and ~$\pi(\chi_n\vert \theta)$
are Gaussian multivariate distributions,
whose mean and covariance are given by the kriging equations~\cite{santner2003design}.

The posterior distribution of~$\theta$
can be expressed with Bayes formula up to a normalizing constant:
$\pi\left(\theta\vert\chi_n\right) \propto
\pi(\chi_n\vert\theta)\cdot\pi(\theta)$.
As there is no close expression of
this posterior distribution, we sample it using a Monte-Carlo method.
More precisely, we use the adaptive Metropolis-Hastings algorithm
proposed by~\cite{haario2001adaptive}
to get samples~$(\theta_j)_{1\leq j\leq p}$,
distributed according to~$\pi(\theta\vert \chi_n)$. 

The sampled hyper-parameters are used to compute
the probability distribution of the \gls{pof}~$P$~\eqref{eq:pof}.
Since the density of~$P$ is intractable,
we use a Monte-Carlo method to draw samples
from the posterior distribution~$\pi(P\vert \chi_n)$.
At each fixed~$\theta_j$,
first, $m$~inputs are drawn according to
the input distribution~$X^{(j)} = \left(x_{i}^{(j)}\right)_{1\leq i\leq m}$,
$x_{i}^{(j)}\sim f_{\ens{X}}$.
Then, we draw $q$~Gaussian sample paths at the inputs~$X^{(j)}$
at the reference level~$\left(\xi_{\chi_n, \theta_j}^{(l)}
\left(x_{i}^{(j)}, \tHF\right)\right)_{1\leq i\leq m,\,1\leq l\leq q}$
and compute the probability function
$p_{j}^{(l)}\left(x_{i}^{(j)}, \tHF\right)
= \Phi\left(\frac{\xi_{\chi_n, \theta_j}^{(l)}\left(x_{i}^{(j)}, \tHF\right) - \zcrit}
{\sqrt{\lambda_{\theta_j}^{(l)}(\tHF)}}\right)$.
Finally, the samples~$(P_{j,l})_{1\leq j\leq p,\, 1\leq l\leq q}$
are computed by averaging on the input space,
$P_{j,l} = \frac{1}{m}\sum_{i = 1}^{m} p_{j}^{(l)}\left(x_{i}^{(j)}, \tHF\right)$.
With this sample, we can estimate the \gls{pof}
with a measure of uncertainty, for instance, by computing the empirical
median and a 95\% confidence interval.

\section{Application}
\label{sec:proba_fail}

The algorithm is illustrated on a random damped harmonic oscillator
from~\cite{au2001estimation}.
Consider~$X(t)$ the solution of the second-order
differential stochastic equation,
driven by a Brownian motion with spectral density equal to one,
and with~$X(t = 0) = 0$ and~$\dot{X}(t = 0) = 0$ as initial conditions.
The parameters of the differential equation,
the natural pulse~$\omega_0$ and the damping ratio~$\zeta$,
are the $d = 2$~inputs of the simulator.
The stochastic equation is solved on a period
$t\in[0; t^{\mathrm{end}}]$, with~$t^{\mathrm{end}} = 30 \mathrm{s}$,
by an explicit exponential Euler scheme,
which approximates~$X$ by a sequence
$\widetilde{X}_n \approx X(n\cdot \delta{t})$.
The time step~$\delta{t}$ is the fidelity parameter.
The multi-fidelity simulator is
\begin{equation}
f: (\omega_0, \zeta, \delta{t})  \mapsto 
\max_{0\leq n\leq \left\lceil\frac{ t^{\mathrm{end}} }{ \delta{t} }\right\rceil}
\left\{\log\left(|\widetilde{X}_n|\right)\right\},
\end{equation}
with~$\omega_0\in [0; 30]~\mathrm{rad s}^{-1}$,
$\zeta \in [0; 1]$ and~$\delta{t} \in [0;1]~\mathrm{s}$.
The cost of this simulator is linear in~$1/\delta{t}$:
observing the level~$\delta{t}$ (in seconds) costs
$C(\delta{t}) = \frac{2.61}{\delta{t}} + 5.45$ (in milliseconds).
The approximate output range is~$r^{\mathrm{out}} = 40$.

\begin{figure}
\begin{center}
\psfrag{exp(0.5*logsig2)}[tc][tc]{$\sigma_0$}
\psfrag{exp(-1*logacc0[1])}[tc][tc]{$\rho^0_1$}
\psfrag{exp(-1*logacc0[2])}[tc][tc]{$\rho^0_2$}
\psfrag{exp(0.5*logESR)}[tc][tc]{$\sqrt{G}$}
\psfrag{exp(-1*logaccE[1])}[tc][tc]{$\rho^{\epsilon}_1$}
\psfrag{exp(-1*logaccE[2])}[tc][tc]{$\rho^{\epsilon}_2$}
\psfrag{exp(1*logdegr)}[tc][tc]{$L$}
\psfrag{exp(0.5*lognoivar[1])}[tc][tc]{$\sqrt{\lambda(\delta{t}_1))}$}
\psfrag{exp(0.5*lognoivar[2])}[tc][tc]{$\sqrt{\lambda(\delta{t}_2))}$}
\psfrag{exp(0.5*lognoivar[3])}[tc][tc]{$\sqrt{\lambda(\delta{t}_3))}$}
\psfrag{exp(0.5*lognoivar[4])}[tc][tc]{$\sqrt{\lambda(\delta{t}_4))}$}
\psfrag{exp(0.5*lognoivar[5])}[tc][tc]{$\sqrt{\lambda(\delta{t}_5))}$}
\psfrag{Prior density}[cl][cl]{Prior density of~$\theta$}
\psfrag{Posterior density}[cl][cl]{Posterior density of~$\theta$}
\includegraphics[width = 0.82\textwidth]{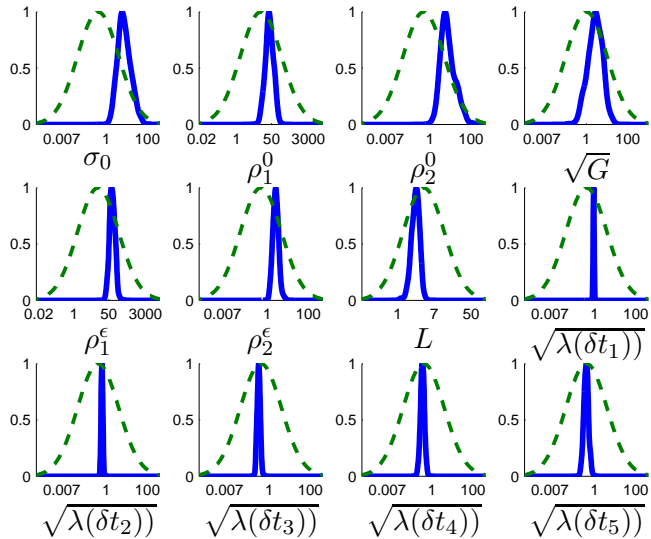}
\end{center}
\caption{Normalized densities of the hyper-parameters
of the multi-fidelity Gaussian process.
The solid blue lines are the posterior densities,
and the green dashed lines the prior densities.
The abscissa axes are in logarithmic scale.}
\label{fig:hyperparameters}
\end{figure}

For this article, we consider~$S = 5$ levels of fidelity:
$\delta{t} = 1$, 0.5, 0.1, 0.05, and 0.01~s.
The multi-fidelity design is a \gls{nlhs} with respectively
168, 56, 28, 14 and 7 points at each level of fidelity,
generated with the algorithm of~\cite{qian2008bayesian} and
a maximin optimization.
The adaptive Metropolis algorithm is applied to draw
$p = 10^3$ vectors~$\theta$.
The figure~\ref{fig:hyperparameters} represents the normalized marginal
prior and posterior distributions of~$\theta$,
the latter being estimated with a kernel density method.

The marginal posterior distributions are more concentrated
than their prior counterparts, indicating that the observations~$\chi_n$ brings
information about the hyper-parameters.
Particularly, for noise variances~$\left(\lambda(\delta{t_s})\right)$,
the strong correlation between levels
allows to reduce the uncertainties of all noise variances,
including those from levels with few observations.
The value of~$L$ is rather well-estimated,
an observation which is opposite to the one in~\cite{tuo2014surrogate},
which recommends to fix the value.

With the sampling of hyper-parameters, we can estimate the posterior
distribution of the \gls{pof}.
The input distribution~$f_{\ens{X}}$ is an uniform distribution on
the input space~$[0; 30] \mathrm{rad} \mathrm{s}^{-1} \times [0; 1]$,
the critical threshold is~$\zcrit = 1$.
In order to make a comparison, we estimate the \gls{pof}
at an observable fidelity-level,
fixed to~$\delta{\tHF} = 0.01 \mathrm{s}$.
We compute a reference value~$P^{\star} = 5.73\%$.

We compare two different methods of estimation: a \gls{fb}
approach, and a plug-in approach, where hyper-parameters are replaced
by their \gls{map}.
Our methodology is applied on 240 independent experiments.
On these experiments, the input and outputs observations change,
but the models and their priors are fixed.
For each experiment, the posterior density of the \gls{pof}
of the four models is sampled,
which gives~$240\times 2$ posterior densities.
From these posterior densities of the \gls{pof}, we compute
the median, and the 95\% confidence intervals.

\begin{figure}
\begin{center}
\psfrag{MF (MAP)}[bc][bc]{}
\psfrag{SL (MAP)}[bc][bc]{}
\psfrag{MF (F-B)}[bc][bc]{}
\psfrag{SL (F-B)}[bc][bc]{}
\psfrag{Estimation on 240 experiments (1; 1)}[tc][tc]
{\acrlong{map}}
\psfrag{Estimation on 240 experiments (2; 1)}[tc][tc]
{\scriptsize{\acrshort{sl}-\acrshort{map}}}
\psfrag{Estimation on 240 experiments (1; 2)}[tc][tc]
{\acrlong{fb}}
\psfrag{Estimation on 240 experiments (2; 2)}[tc][tc]
{\scriptsize{\acrshort{sl}-\acrshort{fb}}}
\psfrag{Density on median (1; 1)}[bc][bc]{}
\psfrag{Density on median (2; 1)}[bc][bc]{}
\psfrag{Density on median (1; 2)}[bc][bc]{}
\psfrag{Density on median (2; 2)}[bc][bc]{}

\subfigure[Posterior medians]
{\includegraphics[width=0.41\textwidth]{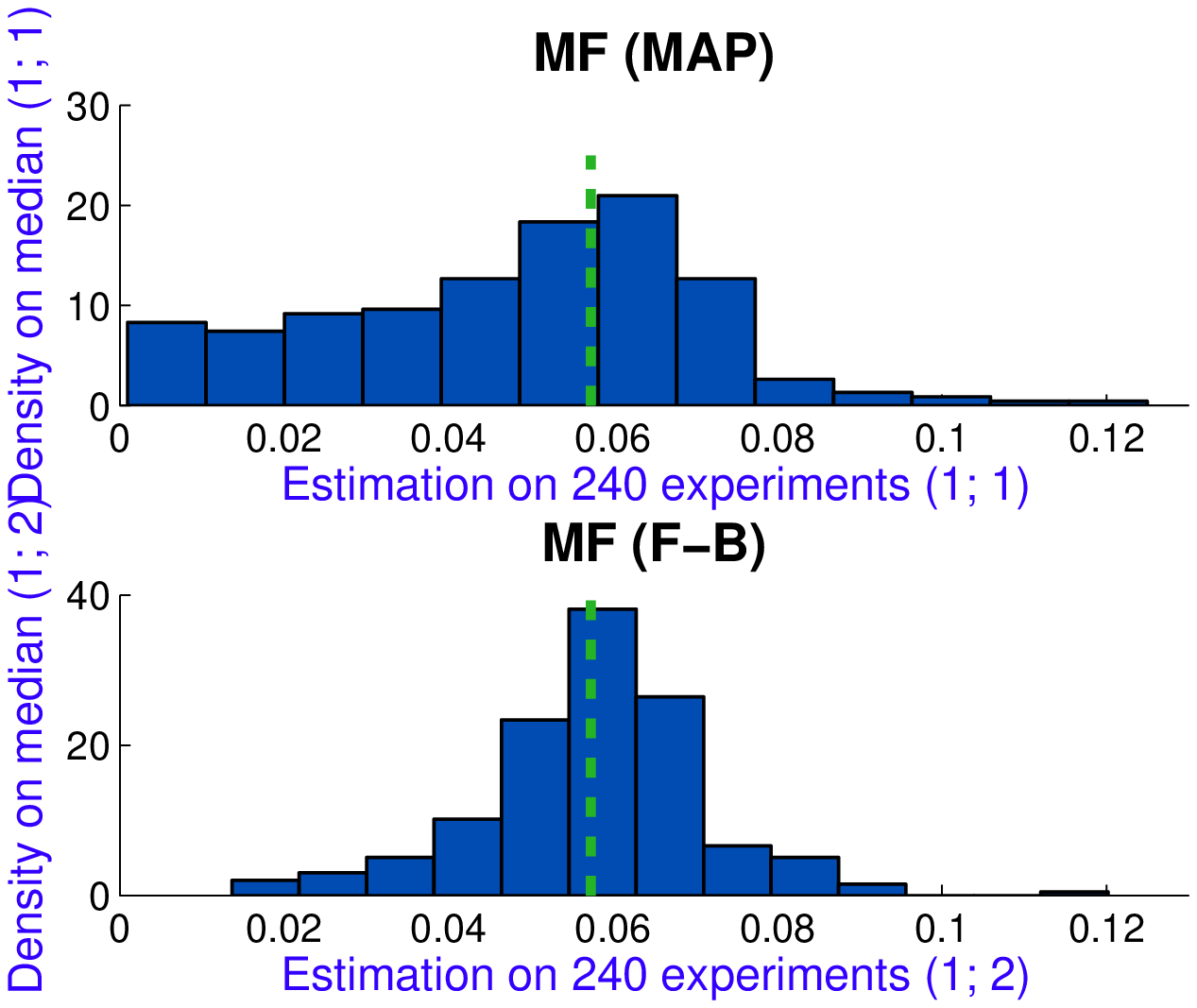}
\label{fig:histogram_results_median}}
\psfrag{MF (MAP) (Success: 179/240)}[bc][bc]{}
\psfrag{SL (MAP) (Success: 120/240)}[bc][bc]{}
\psfrag{MF (F-B) (Success: 230/240)}[bc][bc]{}
\psfrag{SL (F-B) (Success: 182/240)}[bc][bc]{}
\psfrag{Lengths at 95\% (240 experiments) (1; 1)}[tc][tc]
{\acrlong{map}}
\psfrag{Lengths at 95\% (240 experiments) (2; 1)}[tc][tc]
{\scriptsize{\acrshort{sl}-\acrshort{map}}}
\psfrag{Lengths at 95\% (240 experiments) (1; 2)}[tc][tc]
{\acrlong{fb}}
\psfrag{Lengths at 95\% (240 experiments) (2; 2)}[tc][tc]
{\scriptsize{\acrshort{sl}-\acrshort{fb}}}
\psfrag{Density on lengths (1; 1)}[bc][bc]{}
\psfrag{Density on lengths (2; 1)}[bc][bc]{}
\psfrag{Density on lengths (1; 2)}[bc][bc]{}
\psfrag{Density on lengths (2; 2)}[bc][bc]{}
\subfigure[Lengths of 95\% intervals]
{\includegraphics[width=0.41\textwidth]{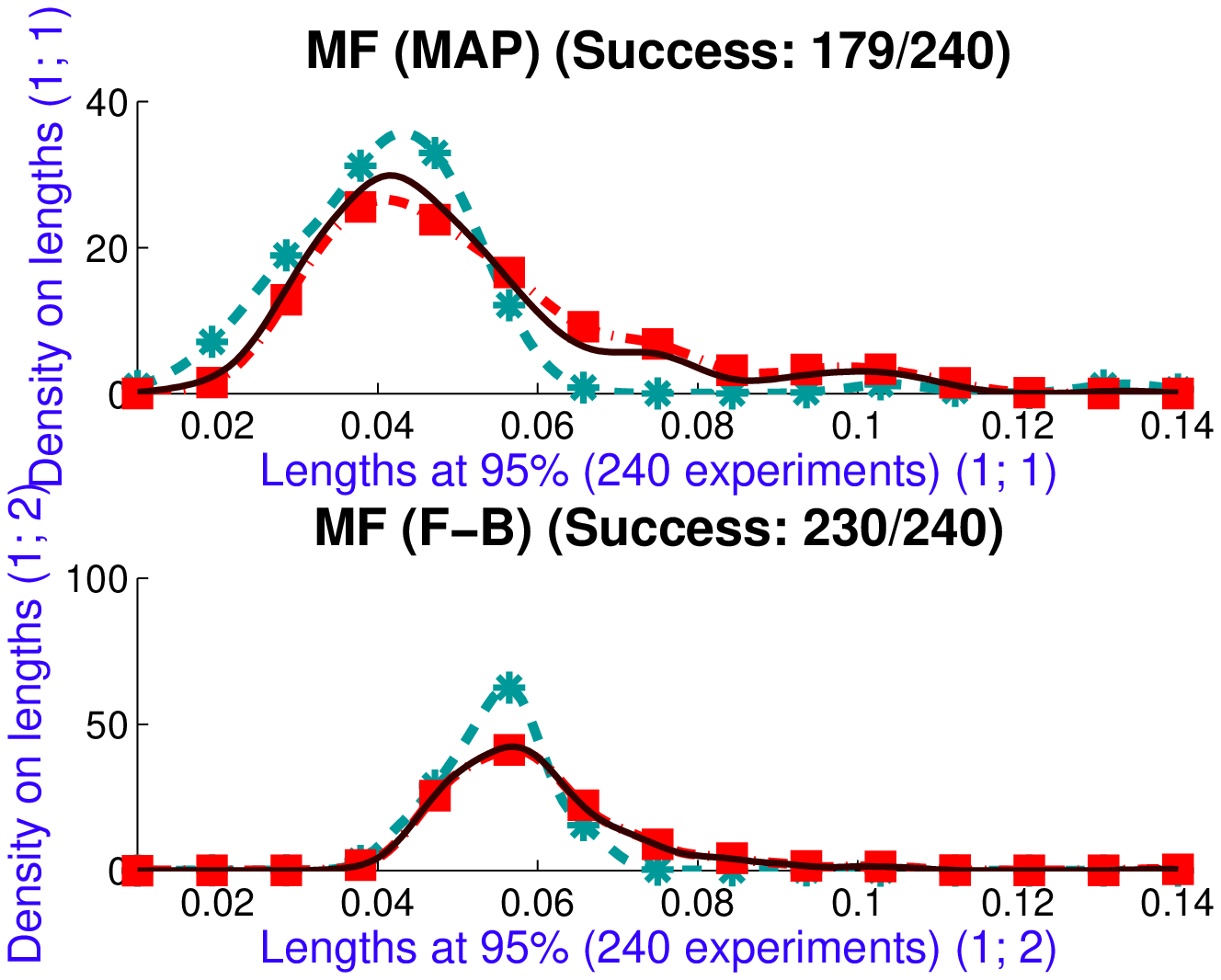}
\label{fig:histogram_results_ci95}}
\psfrag{Probability of the confidence interval}[tc][tc]{Level}
\psfrag{Proportion of catching intervals}[bc][bc]{Coverage}
\psfrag{Coverage (model MF)}[bc][bc]{}
\subfigure[Coverage rate versus level of the confidence intervals]
{\includegraphics[width = 0.41\textwidth]{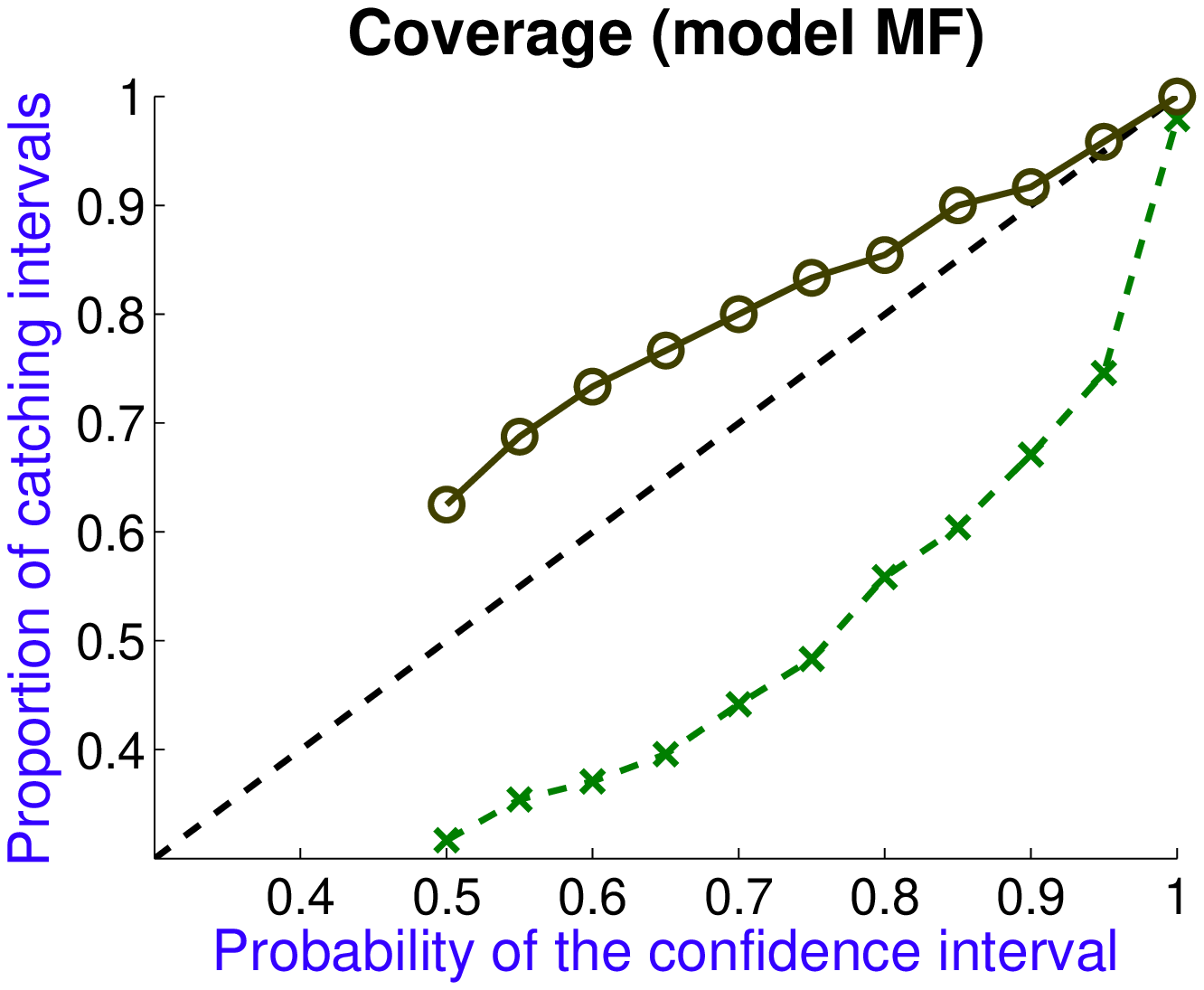}
\label{fig:part_success}}
\end{center}
\caption{(a)~Histograms of the medians.
The vertical dotted line is the reference.
(b)~Estimated densities of the lengths of the 95\% confidence interval.
The dashed lines with squares correspond to intervals
which contain the reference value,
the dashed lines with stars to those which miss it,
and the solid lines to all intervals.
(c)~Coverage of confidence intervals at level~$p$.
The coverage is the proportion of cases
where the reference value is inside the confidence interval.
The solid and dashed-crossed lines corresponds respectively to the \gls{fb}
and \gls{map} approaches.}
\label{fig:histogram_results}
\end{figure}

Figure~\ref{fig:histogram_results_median} displays the empirical
histograms of the 240 medians of the posterior distributions of the \gls{pof}.
We can see that the medians returned by \gls{fb} approach
vary less from an experiment to another
than those returned by \gls{map}.
Figure~\ref{fig:histogram_results_ci95} plots the empirical
densities of the 240 lengths of the 95\% confidence intervals,
estimated by kernel density regression.
We can see that, for all approaches, the failing intervals have a smaller length
than the successful intervals.
We can also see that the \gls{fb} approach always provides
non-zero confidence intervals, opposite to \gls{map} approach.
Figure~\ref{fig:part_success} presents the capacity of the models
to catch the reference value.
Each curve corresponds to one approach.
Each point of the curve at abscissa~$p$ is the coverage,
the proportion of the confidence intervals of level~$p$
which contain the reference, according to the associated approach.
We can see that the \gls{fb} approach provides much more conservative intervals
than \gls{map} approach.

The three Figures~\ref{fig:histogram_results_median}, \ref{fig:histogram_results_ci95}
and~\ref{fig:part_success} suggest that, on this example,
the \gls{fb} approach returns a better posterior distribution
than the \gls{map} approach.

\section{Conclusion}
\label{sec:conclusion}

In this article, we propose a Bayesian model for stochastic
multi-fidelity numerical model.
The model is based on a Gaussian process, completed with
prior distributions on the hyper-parameters of the covariance function
and on noise variances.
By comparing prior and posterior hyper-parameter distributions,
we see that observations bring informations about the hyper-parameters.
Using sampling algorithms, we can sample the posterior
distribution of the quantity of interest, here a \acrfull{pof}.
By comparing the \acrlong{fb} approach with \acrlong{map} plug-in approach,
we can see that, on an academic example, the \acrlong{fb} approach
provides more robust confidence intervals of the \gls{pof}.
However, the priors require care when using the models.
Future work will focus on assessing the impact of the different prior modeling choices
on the posterior distributions of hyper-parameters and of quantities of interest.

\bibliographystyle{plain}
\bibliography{AMCTM_bibliography}

\end{document}